\documentclass[12pt]{article}
\usepackage{graphicx,amssymb,amsmath,amsthm,cite}

\newtheorem{definition}{Definition}

\def\be{\begin{equation}}
\def\ee{\end{equation}}
\def\ben{\begin{eqnarray}}
\def\een{\end{eqnarray}}

\newcommand{\la}{\langle}
\newcommand{\ra}{\rangle}

\newcommand{\til}{\tilde}

\def\Del{\Delta}
\def\tDel{\til{\Delta}}

\def\op{\hat{P}}

\def\SS{{\cal{S}}}

\def\SV{{\cal{V}}}
\def\SW{{\cal{W}}}

\def\SW{{\cal{W}}}

\def\SWC{{\cal{W}^\bot}}

\newcommand{\Spann}{\text{span}}

\def\N{\mathbb{N}}

\def\emptyy{\{0\}}

\title{Subspace modelling for structured noise 
suppression}
\author{Zhiqiang Xu\thanks{This work was initiated during Z. Xu stay at 
 the Mathematics Division of Aston University, UK}\\
Institute of Computational Mathematics\\
Academy of Mathematics and Systems Science\\ 
Chinese Academy of Sciences\\                              
Beijing 100080, China\\ 
\vspace{0.1cm}\\
Laura Rebollo-Neira\\
Mathematics, Aston University\\ 
Birmingham, B4 7ET, UK\\
\vspace{0.1cm}\\
A. Plastino\\
IFLP-CCT-Conicet\\
Universidad National University La Plata\\
CC 727, 1900 La Plata, Argentina}

\begin{document}
\maketitle

\begin{abstract}
The problem of structured noise suppression is addressed by 
i)modelling the subspaces hosting the components of the signal 
conveying the information and ii)applying a non-extensive nonlinear 
technique for effecting the right separation. Although the approach 
is applicable to all situations satisfying the hypothesis of 
the proposed framework, this work is motivated by a particular scenario, namely,
the cancellation of low frequency noise in broadband 
seismic signals.

\vspace{1cm}

\end{abstract}

\section{Introduction}
The problem of structured noise suppression concerns the elimination 
of signal components produced by phenomena interfering with the 
observations of interest. 
This problem can be addressed by linear techniques 
provided that the subspaces hosting the signal components are complementary 
and well separated \cite{BS94,Reb07a,Reb09,RP09}.
More precisely, if a signal represented by the ket $|f\ra$  is produced as 
the superposition of two components $|f_1\ra\in \SS_1$ and  
$|f_2\ra \in \SS_2$,
provided that $\SS_1 \cap \SS_2 = \emptyy$,
the components of the superposition  $|f\ra=|f_1\ra+|f_2\ra$  can be 
separated by an oblique projection.
Even when this condition is 
theoretically fulfilled, if the subspaces $\SS_1$ and $\SS_2$ are not 
well separated, the concomitant linear problem for extracting
one of the signal components may be ill posed, which  
causes the failure to 
correctly split the signal by a 
linear operation.
Hence, nonlinear techniques for determining a subspace $\SV \subset \SS_1$, 
such that $|f_1 \ra \in \SV$, and the projection onto $\SV$ along 
$\SS_2 $ is well posed, have 
been considered \cite{Reb09,RP09,Reb07b}. In those publications 
the theoretically complementary subspaces $\SS_1$ and $\SS_2$ 
are assumed to be known.
Nevertheless, the condition $\SS_1   \cap  \SS_2 = \emptyy$ 
is strong and the possibility of meeting it depends on the 
ability to generate the 
right model for the subspaces. Unfortunately, the modelling of the 
complementary subspaces by pure physical considerations is not always 
possible and one needs to relay on more general mathematical modelling.

Although the technique for subspace modelling we introduce here 
is applicable to different situations, the work is motivated 
by a particular problem relevant to the processing 
of seismic signals. In the nearshore these signals may be 
affected by a population of low-frequency 
waves called {\em{infragravity waves}}\cite{DVP97}. This type
of noise may be also unavoidable in bottom broadband 
seismic observations \cite{DRU07}.
The interested reader is refereed to \cite{CW00} for 
explanations on how  infragravity waves are generated. 
We restrict our consideration to the problem of reducing that type 
of structured low frequency noise from broadband seismic signals. 

Our purpose is twofold. We aim at i)mathematically modelling 
the subspaces to represent
the signal components ii)provide a sparse enough representation of the 
signals so as to make 
sure that the correct splitting can be realized. 

Under the hypothesis that 
one of the signals components lies in the subspace of low frequency 
signals, we determine the subspace of the other component 
in an adaptive manner.
We assume that such a component belongs to an unknown spline space 
and determine the 
knots characterizing the space by taking into account the curvature points 
of the signal in hand. In that sense, the space is `adapted' to 
the particular signal being analyzed. In line with \cite{RP09} 
we tackle the problem of finding  the representation of 
this component through the minimization of the $q$-norm like 
quantity, which is closely related to the non-extensive entropy introduced 
as ingredient of a thermodynamic framework in the seminal paper 
 by Tsallis \cite{Tsa88}
and ever since broadly applied in physics 
\cite{Tsa88,Tsa09,Pla1,Pla2,AO01,AS01,AAM02,ASI04} 
and other disciplines \cite{GT04}.  

The paper is organized as follows: 
In Section 2 we address the problem of subspaces modelling 
 and discuss the non-extensive nonlinear technique yielding the 
  right signal splinting.
A numerical simulation concerning the filtering of 
low frequency noise from a seismic signal is presented 
in Section 3. The conclusions are drawn in Section 4.

\section{Adaptive subspace modelling for structured noise filtering}
As already mentioned, we are concerned with the problem of 
separating from a signal those components which are not relevant 
to the phenomenon of interest. For simplicity we consider that 
a signal $|f\ra$ 
is the superposition of only two components and the goal is to 
find a suitable model for the subspaces hosting such component. 
Since out work is motivated by the specific problem of filtering 
low frequency noise from a seismic signal, we further assume that 
the subspace representing that type of structured noise is spanned by a few Fourier functions.  In accordance with previous works we 
denote such a subspace as $\SWC$ and consider it to be fixed.
The composed signal is the superposition
$|f\ra= |f_\SV \ra + |f_\SWC \ra$  with $|f_\SV \ra \in \SV$. 
The goal is to model 
the subspace $\SV$ fulfilling the theoretical 
condition $\SV \cap \SWC = \emptyy$, regardless of the fact that 
the two subspaces may be too close to each 
other for the signal separation to be obtained via a linear approach. 
We allow for this  difficulty by introducing the additional 
hypothesis that $|f_\SV \ra$ is well represented in a subspace of 
$\SV$, which is tantamount to assuming that $|f_\SV\ra$ has  
a `sparse enough' representation in $\SV$. We further 
assume that $|f_\SV \ra$ is well approximated in a
dedicated spline space to be constructed as described in the 
next section.

\subsection{Finding the appropriate spline space}
Let us start by stating the few definitions on spline 
spaces which are needed for setting up our mathematical framework. 
For a complete treatment of splines we refer to the fundamental 
books \cite{Sch81,Chu88,Boo01}.
\begin{definition}
Given a finite closed interval $[c,d]$ we define
a {\em{partition}} of $[c,d]$ as the finite set of points
\begin{equation}\label{Delta}
\Del:=\{x_i\}_{i=0}^{N+1}, N\in\N,\,\,\text{such that} \,\,
c=x_0<x_1<\cdots<x_{N}<x_{N+1}=d.
\end{equation}
We further define $N$ subintervals $I_i, i=0,\dots,N$ as:
$I_i=[x_i,x_{i+1}), i=0,\dots,N-1$ and $I_N=[x_N,x_{N+1}]$.
\end{definition}
\begin{definition}\label{splinespace}
Let $\Pi_{m}$ be the
space of polynomials  of
degree smaller or equal to $m\in\N_0=\N\cup\{0\}$.
Let $m$ be a positive integer and define
\begin{equation}
S_m(\Del)=\{f\in C^{m-2}[c,d]\ : \ f|_{I_i}\in\Pi_{m-1},
i=0,\dots,N\},
\end{equation}
where  $f|_{I_i}$ indicates the
restriction of the function $f$ on the
interval ${I_i}$.
\end{definition}
An {\em{extended partition}}
with single inner knots associated with $S_m(\Del)$ is a
set $\tDel=\{y_i\}_{i=1}^{2m+N}$ such that
$$y_{m+i}=x_i,\,\,i=1,\ldots,N,\,\, x_1<\cdots<x_N$$
and the first and last $m$  points $y_1\leq \cdots \leq y_{m} \leq
c,\quad d \leq y_{m+N+1}\leq \cdots \leq y_{2m+N}$ can be
arbitrarily chosen.
With each fixed extended partition $\tDel$ there is associated a
unique B-spline basis for $S_m(\Del)$, that we denote as
$\{B_{m,j}\}_{j=1}^{m+N}$. The B-spline $B_{m,j}$ can be defined
by the recursive formulae \cite{Sch81}:
\begin{eqnarray*}
B_{1,j}(x)&=& \begin{cases}
1, & t_j\leq x<t_{j+1},\\
0, & {\rm otherwise,}
\end{cases} \\
B_{m,j}(x) &=&
\frac{x-y_j}{y_{j+m-1}-y_j}B_{m-1,j}(x)+\frac{y_{j+m}-x}{y_{j+m}-y_{j+1}}B_{m-1,j+1}(x).
\end{eqnarray*}
For each order, $m$, the corresponding spline space 
is determined by the number 
and position of the knots.  In Fig~1 we show
B-spline basis for two different 
cubic spline spaces ($m=4$) having the same number of 
knots but located at different positions.

\begin{figure}[!ht]
\begin{center}
\includegraphics[width=8cm]{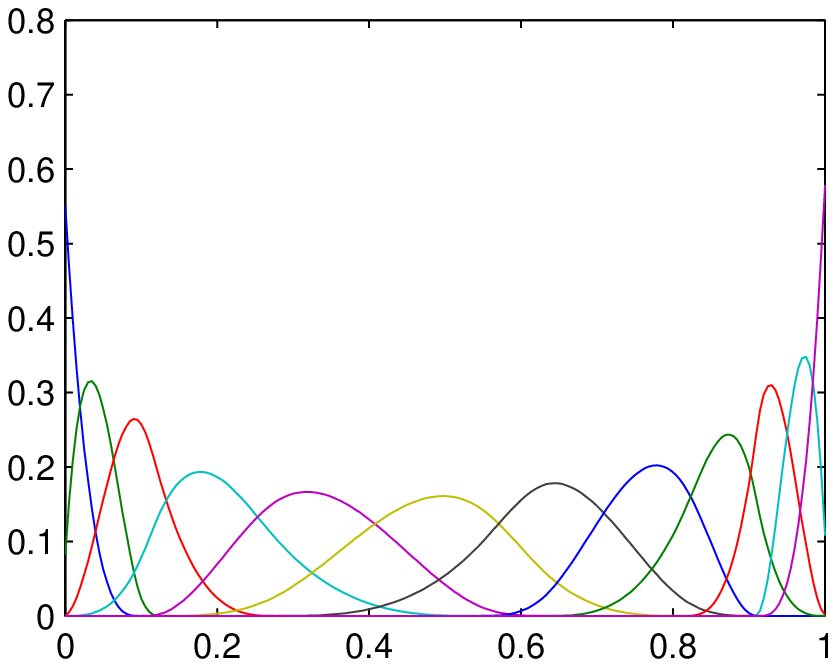}
\includegraphics[width=8cm]{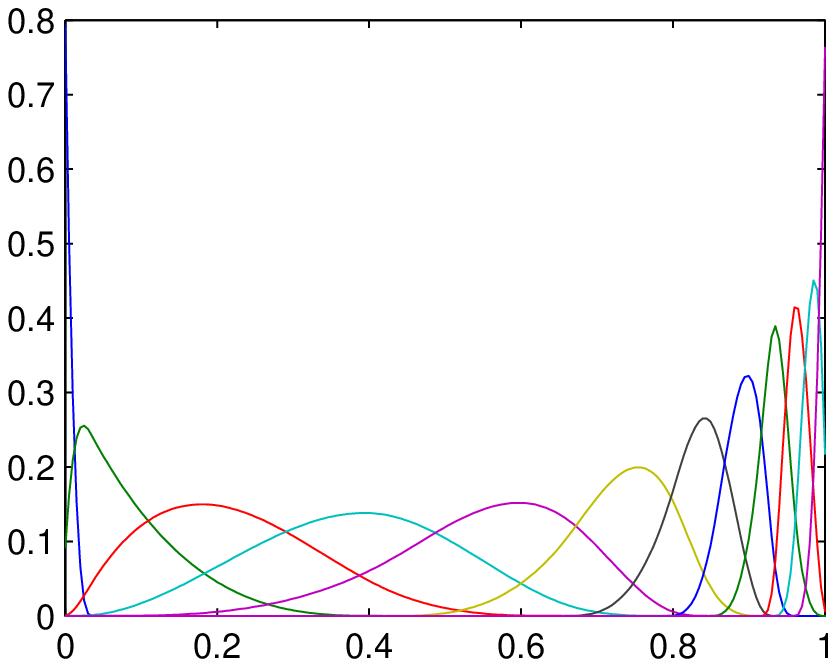}
\end{center}
\vspace{-1cm}
\caption{\small{Left graph: cubic B spline basis functions for an 
arbitrary partition. Right graph: same description as in the left 
graph but for a different partition with the same number of knots}}
\end{figure}
In order to find the appropriate partition $\Del$ giving rise to 
the appropriate spline space $S_m(\Del)$ to represent a given 
signal $f$, we first determine the critical points of
the signal's curvature, i.e., we find the set $T$ 
defined as
\be
\label{cur}
T\,\,:=\,\,\{t: \left(\frac{f''}{(1-f'^2)^{3/2}}\right)'(t)=0\}.
\ee
The entries in $T$ are chosen as the initial knots of
$\Del$. Extra knots are obtained by subdivisions
between consecutive knots in $T$ so as to generate a partition
$\Del$ with the desired number of knots. An algorithm for
implementing this procedure on a signal given  as a  
discrete piece of data is outlined in \cite{RX09}.
We would like to be able to use this procedure on 
the signal we need to represent, namely the component 
$|f_\SV\ra$, but, of course we do not have access to this
signal; our goal is to find it!. Thus, in line with 
\cite{Reb09,RP09} we proceed as explained below. 

Since in our framework ${\SWC}$ is fixed and known, we can 
construct the orthogonal projector onto ${\SW}=(\SWC)^\bot$ that we 
denote $\hat{P}_{\SW}$. Assuming now that $\SV= S_m(\Del)$,
for some order $m$ and some partition $\Del$, we can 
use B-splines to span the space so that for $|f_\SV \ra \in S_m(\Del)$
we have
\be
\label{fsv}
f_\SV(x) = \la x| f_\SV \ra= \sum_{i=1}^M c_{i} \la x |  B_{i,m}\ra=
\sum_{i=1}^M c_{i} B_{i,m}(x).
\ee
Hence, by applying the projector $\hat{P}_{\SW}$  on both sides of
\eqref{fsv} 
we further have
\be
\label{fsu}
|f_\SW \ra = \sum_{i=1}^M   c_{i}  |u_{i}\ra,\quad \text{where}\quad 
|u_{i}\ra= \hat{P}_{\SW} |B_{i,m}\ra ,\quad \text{and}\quad
|f_{\SW}\ra= \hat{P}_{\SW} |f_{\SV}\ra.
\ee
Denoting  by $\hat{I}_{\SS}$ the identity operator in $\SS=\SV + \SWC$, the
projector $\op_{\SW}$ is obtainable from the relation 
$\op_{\SW} = \hat{I}_{\SS}- \op_{\SWC}$. Therefore
 the component $|f_{\SW}\ra$ is available and can be used  to 
determine the knots of the spline space to 
represent it. 

Let us  suppose then that through the curvature 
function \eqref{cur} we obtain a suitable partition for 
the space to represent $|f_{\SW}\ra$. Now, in order to 
 obtain the component $|f_{\SV}\ra$ from $|f\ra$ in the most 
 usual case involving subspaces $\SV$ and $\SWC$ close to 
each other, we need to find the representation of 
$|f_{\SW}\ra$ in a subspace of 
$\SW=\Spann\{|u_i\ra\}_{i=1}^M$. The approach for achieving  such an 
aim is discussed in the next section. 

\subsection{Determination of the signal representation through a 
nonlinear non-extensive approach}
At this point we can assume that we know the spanning 
set for $\SW$ so that henceforth the problem is reduced to 
finding the representation \eqref{fsu} with sparse coefficients.
For this we could apply the approach proposed in \cite{RP09}, 
which entails to use the normal equations 
\be
\label{noreq}
\la u_n| f_W\ra =\sum_{i=1}^M c_i \la u_n | u_i\ra,\quad n=1\ldots,M.
\ee
as constraints for the minimization of the $q-$norm like quantity
$\sum_{i=1}^M |c_i|^q,\,0 < q \le 1$  but incorporating the 
equations in a stepwise manner. However, in the case motivating this work
the number of necessary constraints is large enough to make  
the whole process slow. Hence 
rather than using  the approach proposed in
 \cite{RP09}  we take an alternative route and apply a regularized 
version of the FOCUSS algorithm, which implies to 
minimize the functional 
$${\cal{L}}=\sum_{i=1}^M |c_i|^q + \lambda \||f_\SW \ra - \sum_{i=1}^M c_i  |u_i \ra\|^2,$$
where $\lambda$ is a regularization parameter. 
The algorithm for implementing the approach is based 
on re-weighted least squares and 
is given in \cite{REC03}. It comprises the 
following simple steps
\begin{itemize}
\item[1)] For each fixed $q$
set a value for $\lambda$ and a value for the initial 
vector $|c^o\ra=  \sum_{i=1}^M c_i^o |i \ra$. 

\item[2)]
At each iteration, say iteration $k$, define the operators
$$\hat{B}=\sum_{i=1}^M |B_{i,m}\ra \la i|, \quad \hat{\Pi}_k 
= \sum_{i=1}^M |i\ra |x_i^{k-1}|^{1-\frac{q}{2}} \la i|, 
\quad \hat{B}_k = \hat{B} \hat{\Pi}_k, $$
where the representation $\la x|B_{i,m}\ra\,i=1\ldots,M$ of the kets
$|B_{i,m}\ra\,i=1\ldots,M$ are B-spline 
functions $\la x|B_{i,m}\ra=B_{i,m}(x)$ of order $m$.

\item[3)] Compute $|c^{k}\ra$ as
$$|c^{k}\ra= \hat{\Pi}_k (\hat{B}_k^\ast \hat{B}_k + 
\lambda \hat{I})^{-1} | f_\SW\ra,$$ 
where $\hat{B}_k^\ast$ indicates the adjoint of $\hat{B}_k$ and 
$\hat{I}$  the identity operator. 

\item[4)]Given a small $\epsilon$, while
$\||c^{k}\ra - |c^{k-1}\ra\|>\epsilon$,  repeat 2) and 3)
\end{itemize}
For the derivation of the method and  
discussion on convergence issues see \cite{REC03}. 

When the numerical convergence has been reached, say at iteration $K$, 
set $c_i= c_i^K,\,i=1,\ldots,M$ and compute 
the required component $|f_\SV\ra$ as

$$|f_\SV\ra= \sum_{i=1}^M c_i |B_{i,m}\ra.$$ 

\section{Application to filtering of
structured low frequency noise from a seismic signal}
We apply here the proposed approach to filtering low frequency 
noise from a seismic signal. As already mentioned, a common interference 
with broadband seismic signals is produced by long waves, generated  by 
known or unknown sources, called infragravity waves 
\cite{DVP97}.  This interference is 
referred to as low frequency noise as it falls in a frequency range 
of up to 0.05 Hz. Thus, the model for the subspace of that type of 
structured noise, on a signal given by $L=403$ samples, is
\be
\label{nosub}
\SWC= \Spann\{e^{\imath \frac{2\pi n(i-1)}{L}},\,i=1,\ldots,L\}_{n=-21}^{21}.
\ee
The particular realization of the noise we have simulated 
is plotted in the top left graph of Fig~2
(signal $f^n_i,\,i=1,\ldots,L$). 
However, it is appropriate to recall 
that the success of the approach does not depends on the 
actual form of the noise (as long as it belongs to the subspace
 given in \eqref{nosub}) because the approach guarantees the 
suppression of the whole subspace $\SWC$. 

The seismic signal $f^s_i,\,i=1,\ldots,L$ shown in the right 
graph of Fig~2 is 
a piece of a test signal distributed by the seismic industry.
The left bottom graph is the  superposition of the signals 
in the top graphs, i.e. $f_i= f^n_i + f^s_i,\,i=1,\ldots,L.$
We fist subtract from $f$ the component in $\SWC$ to 
obtain $f_{\SW}= f - \hat{P}_\SWC f$ and use this 
signal to find the points in the set \eqref{cur}. Then 
we subdivide uniformly those points to obtain $341$ nonuniform
knots defining the signal space. Using these knots we
construct the nonuniform B-spline basis for the space 
(MATLAB codes for the implementation of 
both steps are available from \cite{web2}). 
\begin{figure}[!ht]
\begin{center}
\includegraphics[width=7cm]{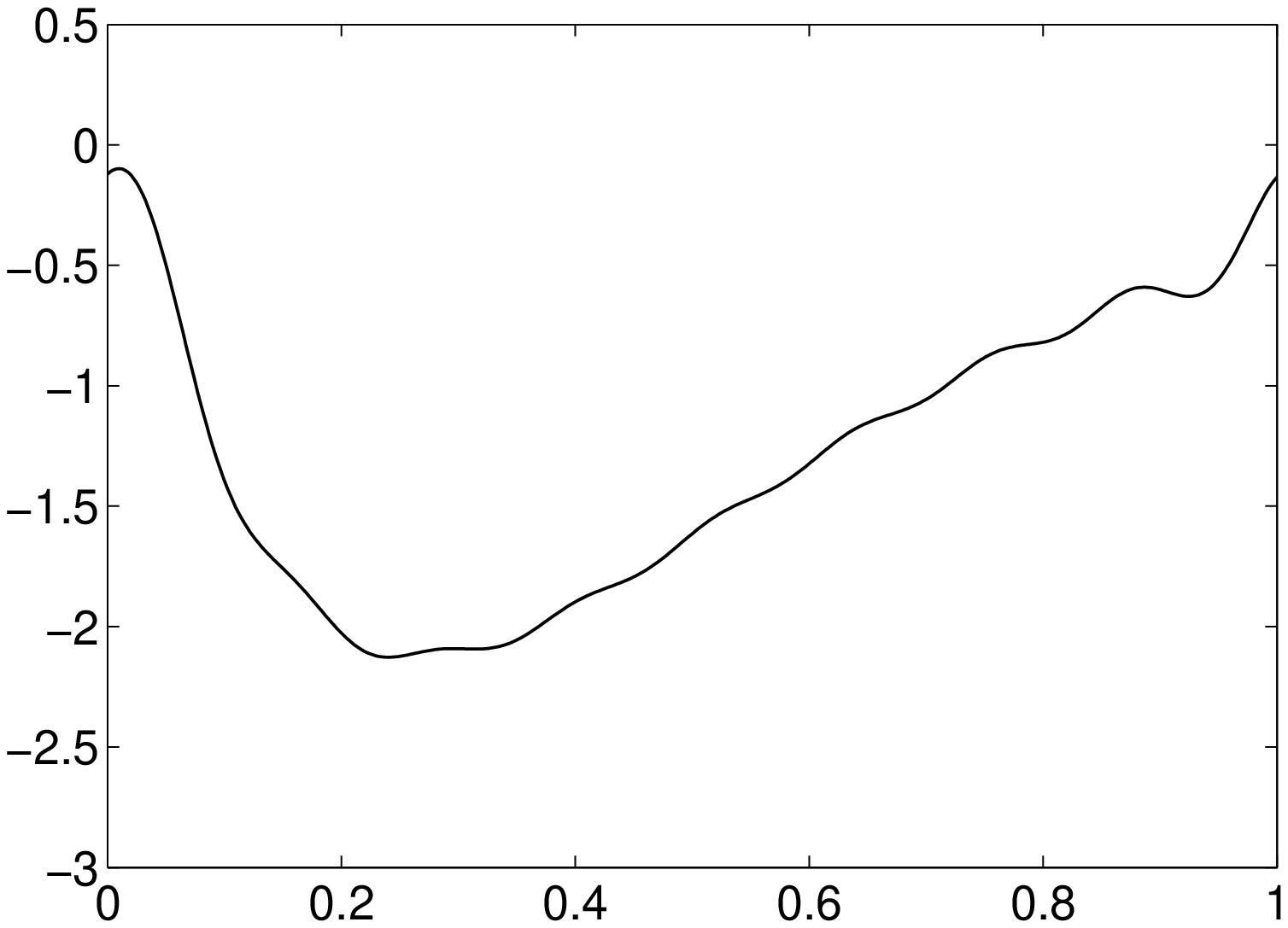}
\includegraphics[width=7cm]{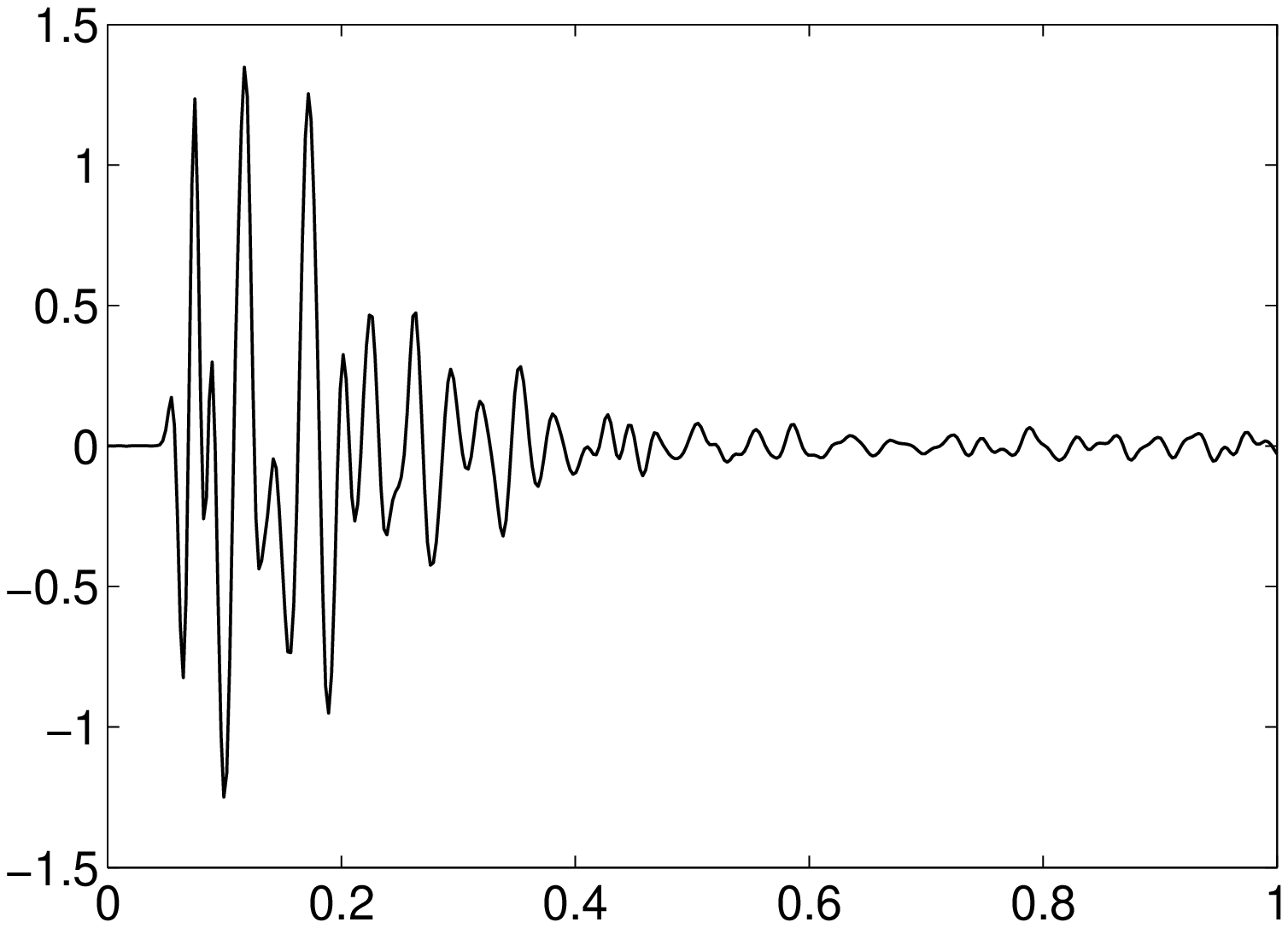}\\
\includegraphics[width=7cm]{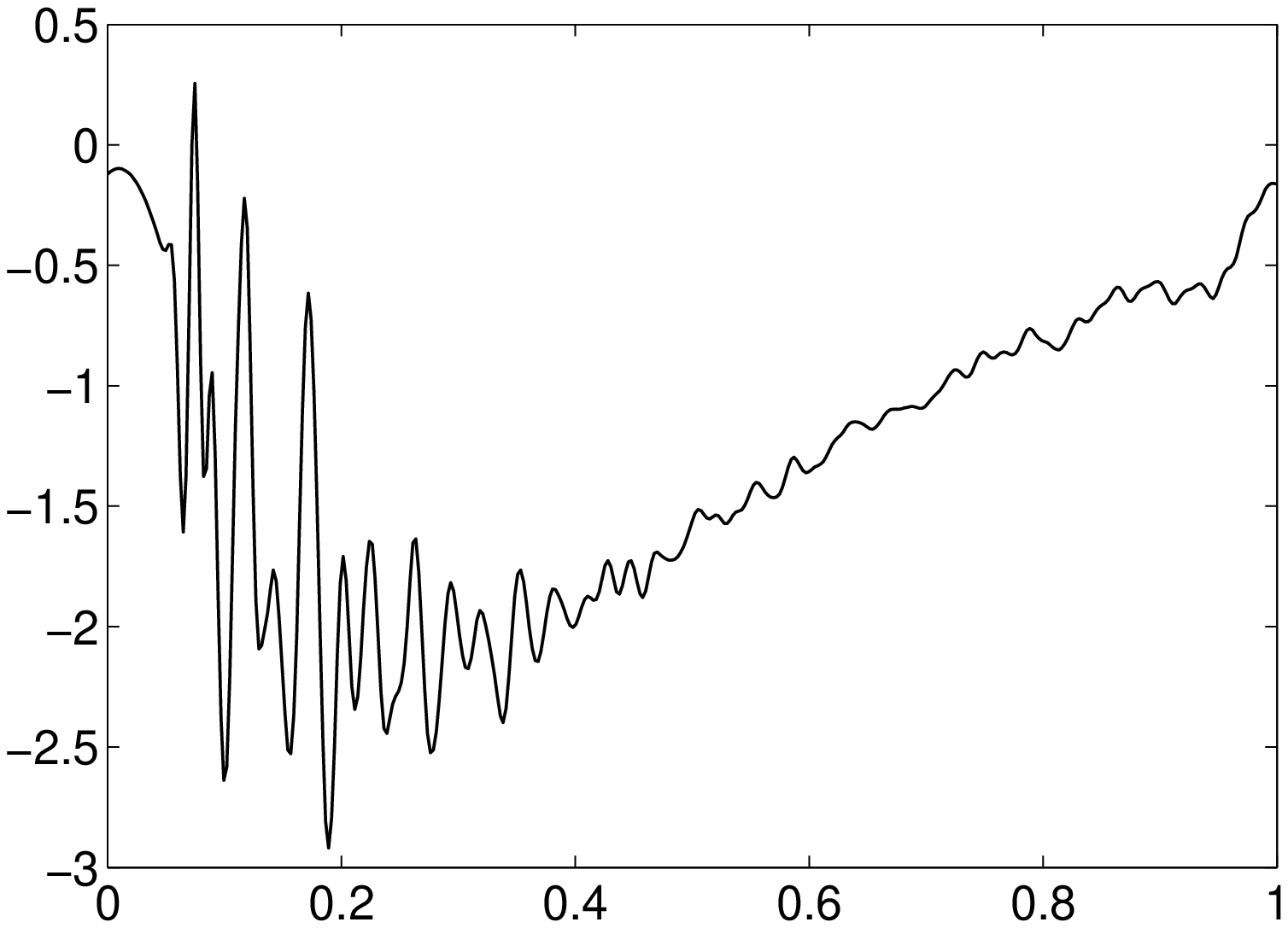}
\includegraphics[width=7cm]{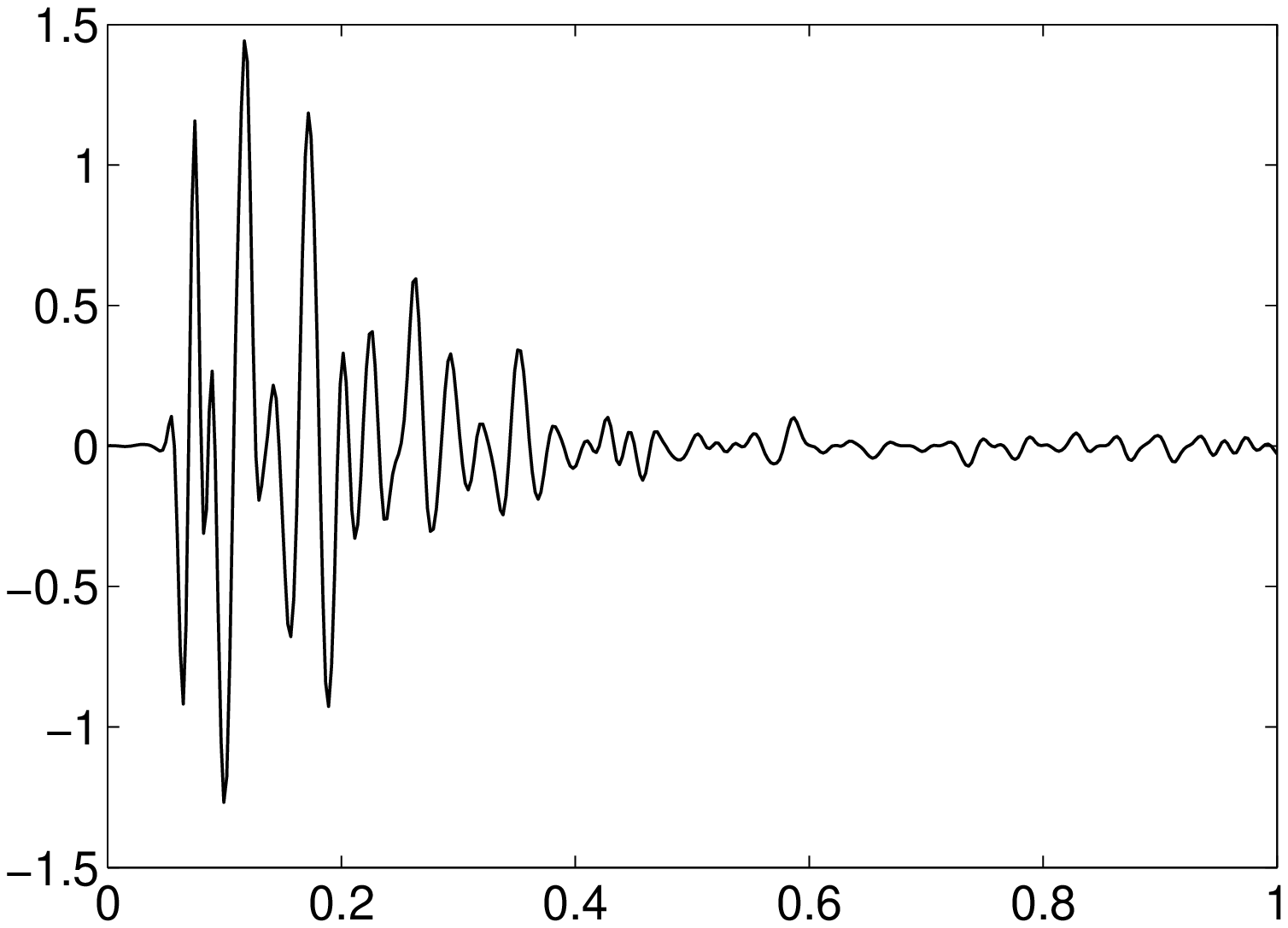}
\end{center}
\vspace{-1cm}
\caption{\small{Top left graph: Simulated low frequency noise $(f^n)$.
Top right graph: piece of seismic signal distributed as test 
signal by the seismic industry $(f^s)$. 
Left bottom graph: Signal plus noise $(f=f^s+f^n)$.
Right bottom graph: Approximation $f^q$ recover from the 
left graph by applying the proposed approach for $q=0.123$.}}
\end{figure}

The methodology discussed in the previous section requires 
to fix the values for $q$  and $\lambda$. 
To allow for good resolution the regularization parameter 
$\lambda$ is given a small value $\lambda= 10^{-8}$. As 
for the parameter $q$ we have let it
vary in the interval $(0,1]$ with step $0.001$ and the best 
value resulted to be $q=0.123$. 
The  right bottom graph of Fig~2 depicts the filtered signal 
arising  by applying the regularized FOCUSS method for 
$\lambda= 10^{-8}$  and $q=0.123$.

The left graph of Fig~3 plots, 
$|f^q- f^s|$, the absolute value of the 
difference between the approximation $f^q$ (for $q=0.123$) and the true 
signal $f^s$. For the sake of comparison, in the 
right graph we have plotted $|f^f- f^s|$, where $f^f$ is 
the approximation arising by filtering with Fast Fourier
Transform (FFT). For this  approximation we simply take 
the FFT of $f$,  eliminate the 
frequencies components in \eqref{nosub}, and apply the inverse
transform to obtain $f^f$. The comparison shows the superiority 
of the proposed approach with respect to standard FFT filtering.

\begin{figure}[!ht]
\begin{center}
\includegraphics[width=8cm]{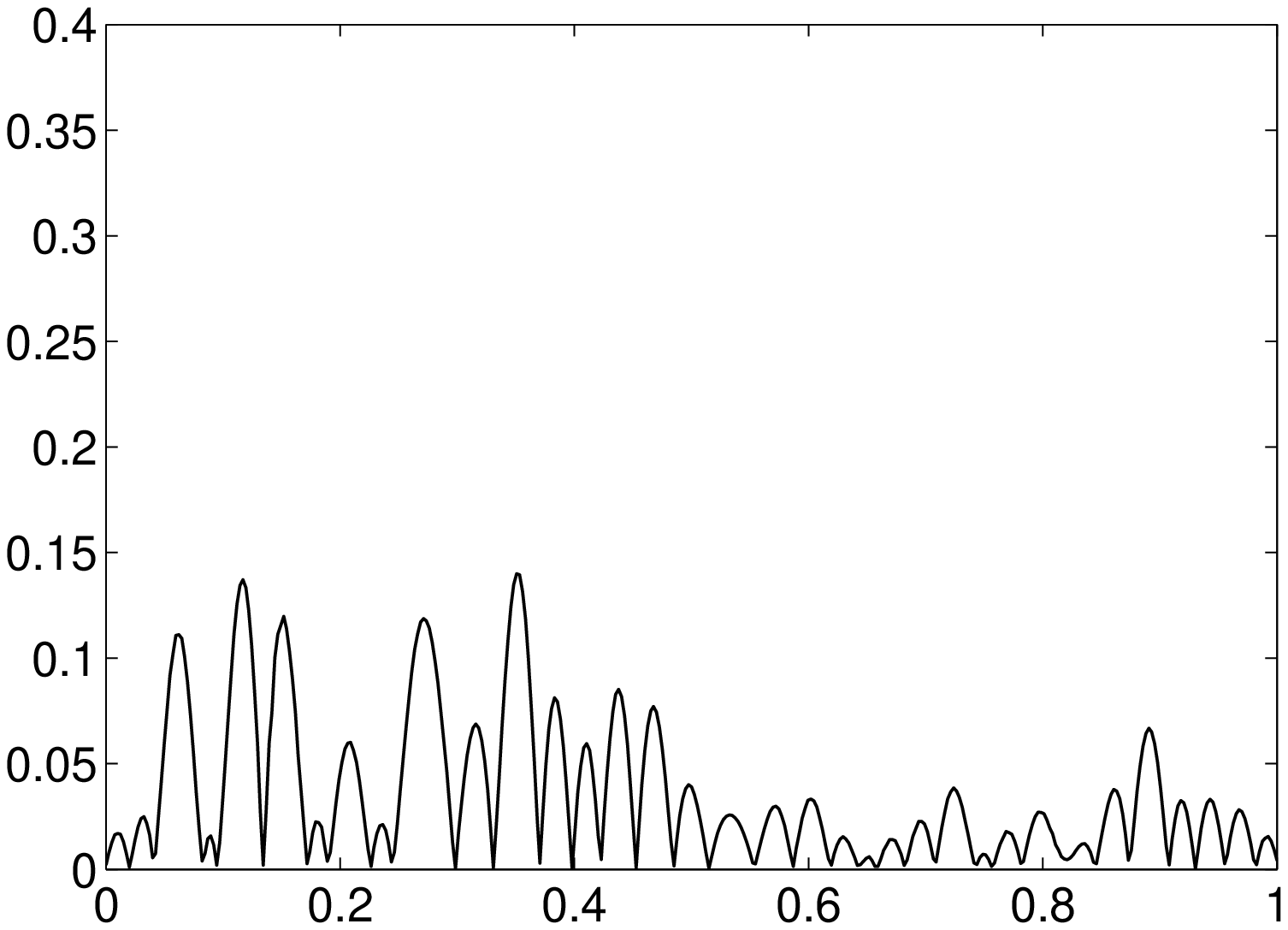}
\includegraphics[width=8cm]{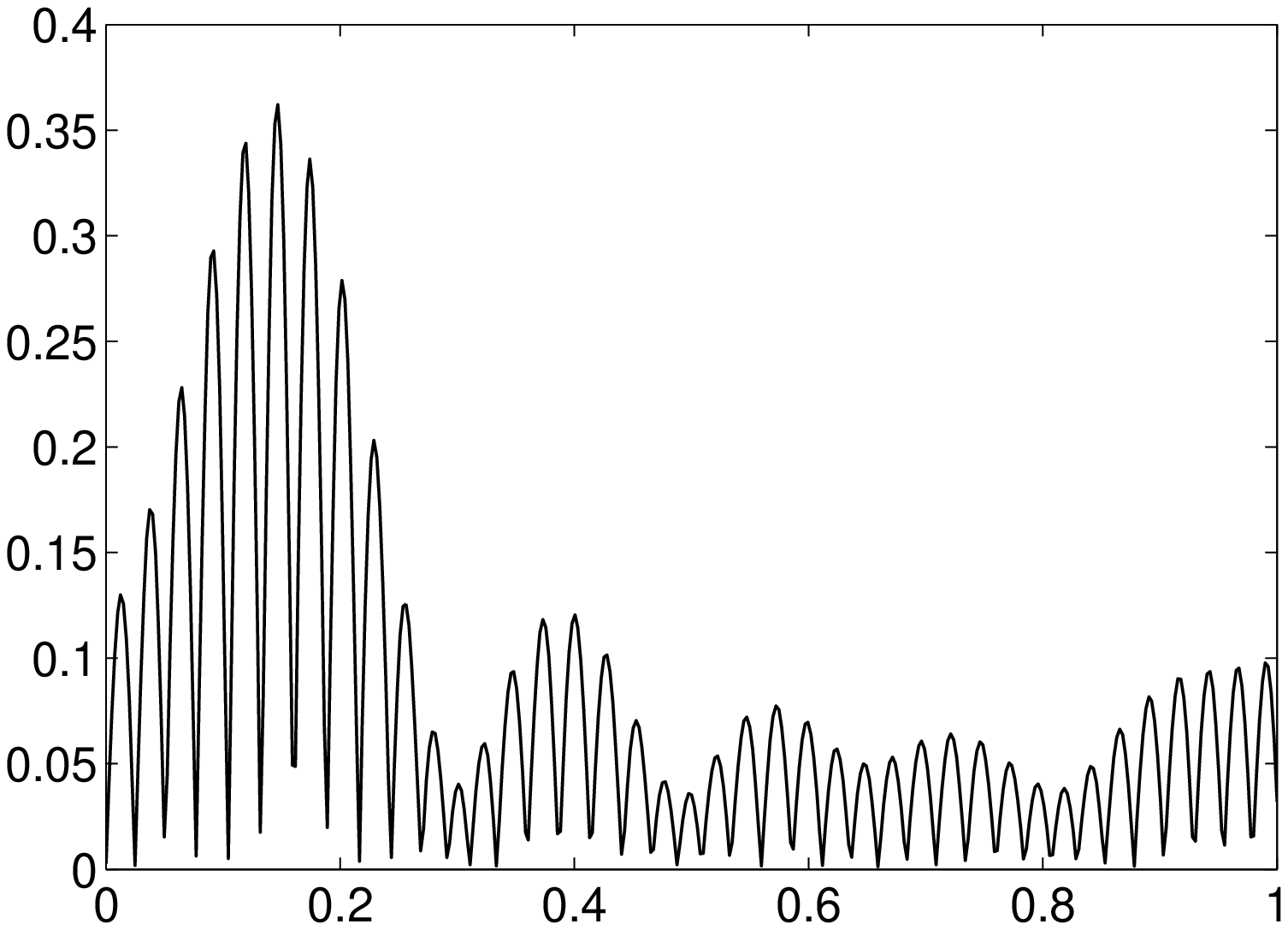}
\end{center}
\vspace{-1cm}
\caption{\small{Left graph: Absolute value of the
difference between the true seismic signal $f^s$ and the
approximation $f^q$ obtained by the proposed approach  
with $q=0.123$. Right graph: Absolute value of the
difference between the true seismic signal $f^s$ and the
approximation $f^f$ obtained by filtering frequencies in the FFT
of $f$.}}
\end{figure}
In order to analyze the dependence of the solution 
on the parameter $q$ 
we calculated the error's norm $\epsilon (q) = \| f^q - f^s\|$. 
The plot of this error, against $q$, is depicted in Fig~4.
While there is clearly an optimum value of 
$q$, the one we have used in this example 
(and a second best value $q=0.312$) 
it should be mentioned that for all the values of 
$q$ in $(0\;1]$ the $f^q$  approximation is superior 
than that obtained by filtering with FFT.
\begin{figure}[!ht]
\begin{center}
\includegraphics[width=8cm]{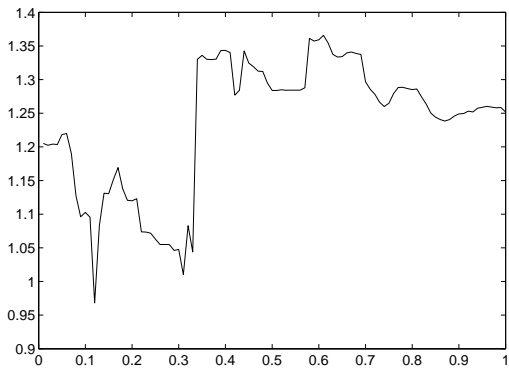}
\end{center}
\vspace{-1cm}
\caption{\small{Norm of the approximation error $\epsilon(q)=||f_s -f_q||$ 
as a function of the parameter $q$.}}
\end{figure}

As a final remark it may be worth stressing that, 
except for $q=1$,
the $q$-norm like quantity is not convex. Nevertheless, 
as opposed to the non-extensive
$q$-entropy \cite{Tsa88,Tsa09} the $q$-norm like quantity is not
extensive for $q=1$.
\section{Conclusions}
The problem of structured noise suppression has been considered by modelling the subspaces of the signal components and applying a nonextensive nonlinear technique for separating them. 
The work was motivated by the problem of filtering
infragravity waves from broadband seismic signals. 
For this, the noise subspace was modelled 
using low frequency Fourier functions and  that  of 
the other component by a dedicated
spline space (adapted to the signal in hand). 
A simulation involving a piece of seismic
signal distributed by the seismic industry 
and noise up to 0.05Hz has produced
encouraging results in comparison with those arising 
by standard Fourier Transform filtering. 
\section*{Acknowledgements}
Support from the Engineering and Physical Sciences Research Council (EPSRC),
UK, grant EP$/$D06263$/$1, is acknowledged.
\newpage
\bibliographystyle{elsart-num}
\bibliography{revbib}

\begin{thebibliography}{10}
\expandafter\ifx\csname url\endcsname\relax
  \def\url#1{\texttt{#1}}\fi
\expandafter\ifx\csname urlprefix\endcsname\relax\def\urlprefix{URL }\fi

\bibitem{BS94}
R.~Behrens, L.~Scharf, Signal processing applications of oblique projection
  operators, IEEE Transactions on Signal Processing 42 (1994) 1413--1424.

\bibitem{Reb07a}
L.~Rebollo-Neira, Constructive updating/downdating of oblique projectors: a
  generalization of the Gram--Schmidt process, Journal of Physics A:
  Mathematical and Theoretical 40 (2007) 6381--6394.

\bibitem{Reb09}
L.~Rebollo-Neira, Measurements design and phenomena discrimination,
J. Phys. A: Math. Theor. 42 (2009) 165210.

\bibitem{RP09}
L.~Rebollo-Neira, A.~ Plastino, Nonlinear non-extensive approach for identification of structured information,
Phyica A, in press (2009)

\bibitem{Reb07b}
L.~Rebollo-Neira, Oblique matching pursuit, IEEE Signal Processing Letters
14~(10) (2007) 703--706.

\bibitem{DVP97}R. D. Kosʹi︠a︡n, N. V. B. Pykhov, B. L. Edge,
Coastal processes in tideless seas, ASCE Publications, 2000.

\bibitem{DRU07} 
D. Dolenc, B. Romanowicz, B. Uhrhammer, P. McGill, D. Neuhauser, 
D. Stakes,
Identifying and removing noise from the Monterey ocean bottom broadband seismic station (MOBB) data,
Geochem. Geophys. Geosyst., 8, (2007) Q02005, doi:10.1029/2006GC001403.

\bibitem{CW00}
W. C. Crawford, S. C. Webb, 
Identifying and removing tilt noise from low-frequency ($<0.1$ Hz) seafloor vertical seismic data, Bull. Seism. Soc. Am., 90, 952-963, 2000. 

\bibitem{Tsa88}
C. Tsallis, Possible generalization of Boltzmann-Gibbs statistics,
J. Stat. Phys., 52, (1988) 479.

\bibitem{Tsa09} C. Tsallis
Introduction to nonextensive statistical mechanics, 
Springer-Verlag, NY, (2009).


\bibitem{Pla1} A. R. Plastino, A. Plastino,  Tsallis
Stellar Polytropes and Tsallis' entropy, Physics Letters A,  174
(1993) 834--386.

\bibitem{Pla2}A. R. Plastino, A. Plastino,  Tsallis Entropy,
Erhenfest Theorem and Information Theory, Physics Letters A,  
177  (1993) 177--179.

\bibitem{AO01} S. Abe, Y. Okamoto, 
Nonextensive Statistical Mechanics and Its Applications
Series: Lecture Notes in Physics , Vol. 560
Springer-Verlag, NY (2001).

\bibitem{AS01}
A, Taruya, 
M. Sakagami,
Gravothermal catastrophe and Tsallis’ generalized entropy of self-gravitating systems,  Physica A, 307, 1-2 (2002) 185--206. 
AS01


\bibitem{AAM02}J. Andrade Jr., M.P. Almeida, A.A. Moreira,
G.A. Farias, Extended phase-space
dynamics for the generalized nonextensive thermostatistics, Phys. Rev. E 65, (2002) 036121

\bibitem{ASI04} G. Adesso, A. Serafini, F. Illuminati, Extremal entanglement and mixedness in
continuously variable systems, Phys. Rev. A 70 (2004) 022318.
`
\bibitem{GT04} M. Gell-Mann, C. Tsallis, (Editors) Nonextensive entropy 
Interdisciplinary Application 
(Santa Fe Institute Studies on the Sciences of Complexity)
Oxford University Press, USA (2004).

\bibitem{Sch81}
L.~Schumaker, Spline Functions: Basic Theory, Wiley, New York, 1981.

\bibitem{Chu88}C. K. Chui,  Multivariate splines, SIAM,
 Philadelphia, 1988.

\bibitem{Boo01}Carl De Boor, A Practical Guide to Splines, Springer,
New York, 2001.

\bibitem{RX09}L. Rebollo-Neira, Z. Xu,
Adaptive non-uniform B-spline dictionaries on a compact interval,
arXiv:0908.0691v1 [math.FA]

\bibitem{REC03} B.D. Rao, K. Engan, 
S. F. Cotter,  J. Palmer, K. Kreutz-Delgado, 
Subset selection in noise based on diversity measure minimization,
IEEE Transactions on Signal Processing, 
 51, 3  (2003) 760-- 770,  10.1109/TSP.2002.808076.

\bibitem{web2} 
http://www.ncrg.aston.ac.uk/Projects/HNLApprox/sigrep2.html 
\end{thebibliography}
\end{document}